\documentclass[11pt]{article}

\usepackage[utf8]{inputenc}
\usepackage[T1]{fontenc}
\usepackage{helvet}
\usepackage{setspace}
\usepackage{verbatim, authblk}
\usepackage{ulem}
\usepackage{graphicx} 
\usepackage{textcomp}
\usepackage{array}
\usepackage{makecell}
\usepackage[a4paper, top=2cm,bottom=2cm,left=0.7in,right=0.7in,marginparwidth=1.75cm]{geometry}
\usepackage{xcolor}
\usepackage{amsmath,amssymb}
\usepackage{upgreek}
\usepackage{float}
\usepackage{stmaryrd}
\usepackage{comment}
\usepackage{relsize}
\usepackage{gensymb}
\usepackage{blindtext}
\usepackage{hyperref}

\usepackage{xr}

\makeatletter
\newcommand*{\addFileDependency}[1]{
\typeout{(#1)}
\@addtofilelist{#1}
\IfFileExists{#1}{}{\typeout{No file #1.}}
}\makeatother

\usepackage[super,numbers,sort]{natbib}

\newcommand{\nab}{\boldsymbol{\nabla}}
\DeclareMathAlphabet\mathbfcal{OMS}{cmsy}{b}{n}

\title{Tunable membrane-less dielectrophoretic microseparation by crossing interdigitated electrodes}

\author[,1]{Nicolas Ruyssen}
\author[2]{Bastien Oliva}
\author[2]{Lylian Challier}
\author[2]{Vincent Noël}
\author[1]{Benjamin Rotenberg \thanks{Corresponding author: \texttt{benjamin.rotenberg@sorbonne-universite.fr}}}
\affil[1]{Sorbonne Université, Physico-Chimie des Electrolytes et Nanosystèmes Interfaciaux, PHENIX, CNRS UMR 8234, Paris F-75005, France}
\affil[2]{Université Paris Cité, Laboratoire ITODYS, CNRS UMR, Paris 7086, France}

\date{}

\begin{document}
\raggedbottom
\maketitle

\paragraph{Keywords}
Dielectrophoresis, Interdigitated electrodes,  Crossed electrodes, Virtual pillars, Membraneless microseparation

\bigskip

\begin{abstract}
Separation is a crucial step in the analysis of living microparticles. In particular, the selective microseparation of phytoplankton by size and shape remains an open problem, even though these criteria are essential for their gender and/or species identification. However, microseparation devices necessitate physical membranes which complicate their fabrication, reduce the sample flow rate and can cause unwanted particle clogging. Recent advances in microfabrication such as High Precision Capillary Printing allow to rapidly build electrode patterns over wide areas. In this study, we introduce a new concept of membrane-less dielectrophoretic (DEP) microseparation suitable for large scale microfabrication processes. The proposed design involves two pairs of interdigitated electrodes at the top and the bottom of a microfluidic channel. We use finite-element calculations to analyse how the DEP force field throughout the channel, as well as the resulting trajectories of particles depend on the geometry of the system, on the physical properties of the particles and suspending medium and on the imposed voltage and flow rates. We numerically show that in the negative DEP regime, particles are focused in the channel mid-planes and that virtual pillars array leads either to their trapping at specific stagnation points, or to their focusing along specific lines, depending on their dielectrophoretic mobility. Simulations allow to understand how particles can be captured and to quantify the particle separation conditions by introducing a critical dielectrophoretic mobility. We further illustrate the principle of membrane-less dielectrophoretic microseparation using the proposed setup, by considering the separation of a binary mixture of polystyrene particles with different diameters, and validate it experimentally. 
\end{abstract}

\newpage

\section{Introduction}

The separation of colloidal particles with typical sizes in the range from one to several tens of micrometer plays an important role in many contexts such as food/beverage industry \cite{el2011cross}, waste-water cleaning \cite{zhang2015optimization}, metal recycling \cite{wang2020continuous,Pesch_review} or the analysis of biological samples \cite{pohl1966separation,Pommer2008,valencia2022direct}. Even though particle microseparation has been successfully achieved in many of these examples, some remain very difficult to perform. In particular, selective microseparation of phytoplankton species from a marine sample is still a great challenge due to the large sample volume required (typically a collection of $10$-$100$~mL of pre-treated samples, resulting from centrifugation, sedimentation, or filtration after Lugol or formaldehyde fixation, of even larger volumes of dilute samples) and the high variability in particle size and shape in the mixture \cite{plankton_2,plankton_1,challier2021printed}. In this context, classical microseparation techniques involving passive physical membranes such as micropillar arrays \cite{alvankarian2013pillar,geng2013continuous,rahmanian2023micropillar} or porous media \cite{yu2022polyphenol} partially loose relevance due to their high hydraulic resistance and the fouling they induce \cite{dizge2011influence}. In addition, passive physical membranes selectively trap particles by size only, which can be a limitation for the discrimination of particles having different shapes but similar sizes. To address this issue of selectivity, it is possible to add an external dielectrophoretic (DEP) force field that attracts particles towards an insulating targeted membrane even at high throughtput \cite{pesch2018bridging} (near $10$~ml.min$^{-1}$) as proposed in Refs.~\citenum{suehiro2003dielectrophoretic,lorenz2020high,pesch2018bridging}. However, this strategy, called insulator-based DEP (iDEP), requires large voltages (from hundreds to thousands of volts), thereby limiting in practice the accessible frequency range below the MHz. Such high frequencies are nevertheless necessary to deal with sizes and compositions typical of biological particles or microplastics.

To overcome this limitation, several membraneless techniques can be found in the literature, to separate microparticles by size \cite{Kim2009}, inertial \cite{warkiani2015membrane,esan2023continuous}, acoustic \cite{collins2014particle,Ma2024} or electric properties \cite{muratore2012biomarker,collins2014particle,kazemi2018numerical,Demierre2007,Wang_dual,WAHEED2018133,Pommer2008}. Their principle is mainly based on controlling the lateral displacement of particles in a fluid flow using an external force field (gravity, electrode based DEP (eDEP), pressure gradient...).  Even though very reliable, these membraneless approaches become cumbersome to perform when the sample includes many types of particles, because they require a number of channel ramifications equal to the number of species to be separated. Another widely used approach is to attract and immobilise particles in the vicinity or at the surface of  electrodes located on the top and/or bottom faces of a microfluidic channel \cite{Rozitsky2013,Becker1995,MORGAN1999516}. 
Particles can be trapped at specific locations, depending not only on their DEP mobility, but also on their initial position in the channel (which is difficult to control experimentally)~\cite{Rozitsky2013,challier2021printed}. Devices able to selectively stop particles within a microfluidic flow at controlled locations, acting as a virtual membrane would enable the high throughput separation and further analysis of large volumes. The principle of such systems would be close to that of optical tweezer arrays \cite{Korda_tweezers,Jenny_tweezers}, but at frequencies much smaller than optical ones, and tunable in order to selectively manipulate particles according to their frequency-dependent dielectric response. However the design and fabrication of such systems remains a great challenge, since one needs to shape the DEP force field across the channel with electrodes that are located at the boundaries. In this context, the ability to manipulate particles using DEP offers in principle an attractive option to achieve this goal.

Recent work using inkjet printing on flexible polymer substrate allows to fabricate simple electrode patterns on top and bottom faces of very wide microfluidic chips, possibly leading to high throughput ($\sim 150$ µl.min$^{-1}$) applications \cite{challier2021printed}. However, inkjet technology has reached its limits for microseparation applications from a prototyping standpoint. Although printers such as the Dimatix (Fujifilm) have cartridges capable of $2.4$~pl droplet production (minimum achievable resolution is $20$~µm to $40$~µm, depending on ink and substrate), nozzles can clog, particularly with inks containing suspended colloids like silver nanoparticles, requiring regular maintenance and possibly production downtime. Partial clogging can also deflect ejected droplet trajectories or create satellite droplets, producing an unusable electrode pattern. For rapid prototyping, these drawbacks present significant difficulties. Compared to inkjet printing, High Precision Capillary Printing (Hummink) offers higher precision and resolution and does not suffer from satellite drops, splashes or drop misalignment. Although the print speed is an order of magnitude lower than that of inkjet, a wider range of ink viscosities and surface tensions is available, so the ink formulation can be optimised for slow drying, limiting the potential for clogging. For rapid prototyping, all of these benefits are highly relevant and seem very promising for microseparation and observation of large volumes of mixture. 

In this study, we propose a new concept of membrane-less dielectrophoretic microseparation adapted to large scale microfabrication techniques such as High Precision Capillary Printing. Such device can be easily fabricated and allows to isolate particles by dielectrophoretic mobility without particle-wall or particle-electrode contact, thereby limiting undesirable mechanotransduction and Joule heating. The concept of the device and its fabrication are presented in Section~\ref{sec:SystemsMethods}, together with the numerical modelling. Section~\ref{sec:ResultsDiscussion} then reports the main results on the dielectrophoretic force field throughout the microfluidic channel and its application to contact-less particle capture, and how they depend on the device geometry, voltage and flow rate. Importantly, we illustrate the principle of membrane-less dielectrophoretic microseparation using the proposed setup, by considering the separation of a binary mixture of PS particles with different diameters, and provide an experimental validation in Section~\ref{sec:experiments}.

\section{Systems and methods}
\label{sec:SystemsMethods}

\subsection{Proposed design and dielectrophoretic driving of particles
}
\label{sect_model}
\begin{figure}[ht!]
     \centering
     \includegraphics[width = \textwidth]{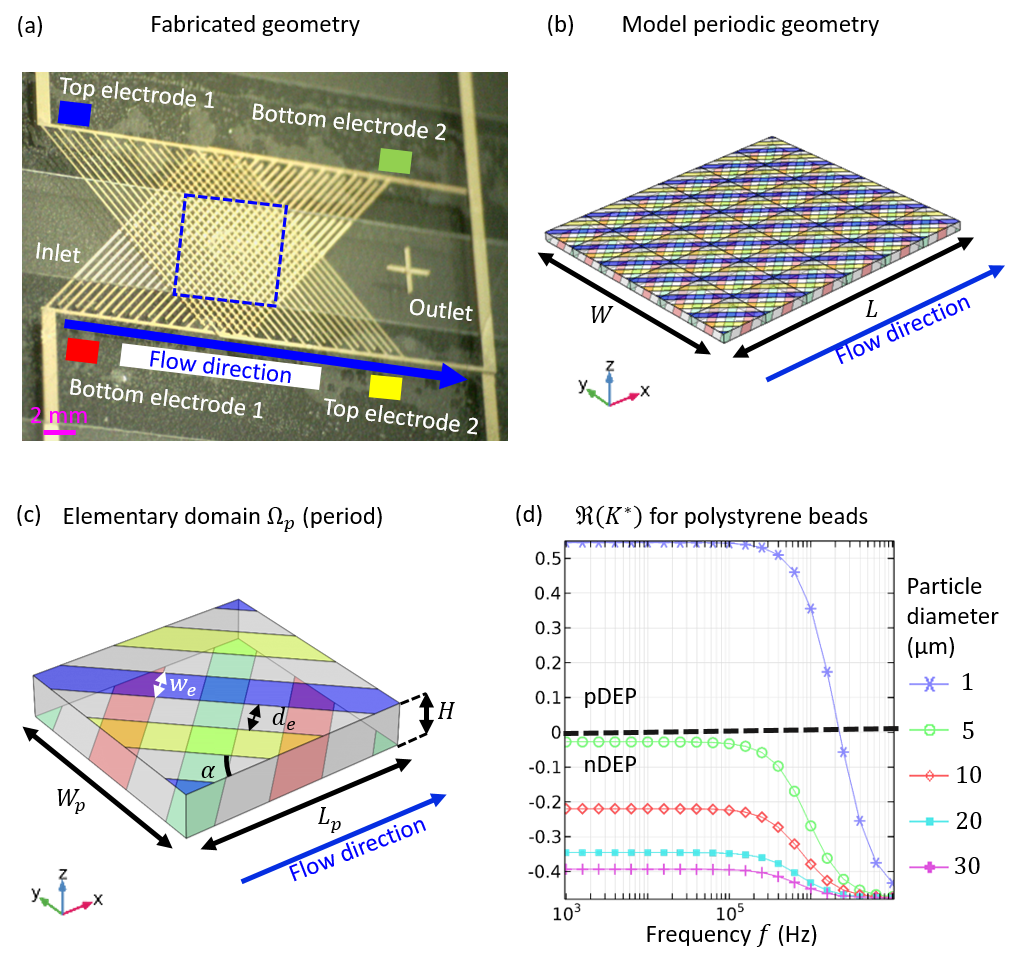}
     \caption{The crossed interdigitated electrodes device's geometry and its $6$ independent parameters: $W,L,H,\alpha,w_e,d_e$. (a) Experimental device; the top and bottom arrays of electrodes are fabricated by High Precision Capillary Printing (see text for details). (b) Periodic representation of the device used for numerical simulations, corresponding the blue boxed region in panel~(a). (c) Elementary domain $\Omega_p$ (period) of the geometry. (d) Frequency dependence (logarithmic scale) of the real part of the Clausius-Mossotti factor $\Re(K^*)$ (linear scale) for polystyrene beads with various diameters $D_p$. The positive (resp. negative) $\Re(K^*)$ domain corresponds to the positive (resp. negative) DEP (pDEP ,resp. nDEP) regime.}
     \label{fig_geom}
 \end{figure}

Figure~\ref{fig_geom} shows the proposed setup (a) and its parametrized periodic representation used for simulations Panels~\ref{fig_geom}(b),(c). The geometry consists in a wide and shallow microfluidic channel with a rectangular cross section of width $W$ (Figure~\ref{fig_geom}(b)) and height $H$ (Figure~\ref{fig_geom}(c)) such that $H\ll W$ (so-called Hele-Shaw cell). The top and bottom faces of the channel each carry two interdigitated electrode arrays, indicated by yellow/blue (resp. red/green) colors for the top (resp. bottom) electrodes in Panels~\ref{fig_geom}(a,b,c). Each electrode array has a fixed electrode width $w_e$ and spacing $d_e$ and both are symmetrically slanted with respect to the channel flow direction by angles $\alpha$ and $-\alpha$, respectively (Figure~\ref{fig_geom}(c)). We note $L$ the length (in the flow direction) of the domain where top and bottom electrode arrays are crossing each other in top view. The resulting geometry is periodic and an elementary domain $\Omega_p$ is shown Figure~\ref{fig_geom}(c); we denote its frontier as $\partial \Omega_p $. The width $W_p$ and the length $L_p$ of $\Omega_p$ are respectively $W_p = 2 \dfrac{w_e+d_e}{\cos(\alpha)}$ and $L_p = 2 \dfrac{w_e+d_e}{\sin(\alpha)}$.

We apply a uniform sinusoidal voltage of Root-Mean-Square (RMS) value $V$ and frequency $f=\omega/2\pi$  between the positively charged electrodes represented in green and yellow Figure~\ref{fig_geom}(a) (connected on the "+" side) and the grounded electrodes indicated in red and blue Figure~\ref{fig_geom}(a). This induces a non-uniform sinusoidal electric potential $\varphi(\mathbf{r},t)  = \sqrt{2} \phi(\mathbf{r})  \cos(\omega t) $ and electric field $\mathbfcal{E}(\mathbf{r},t)  = \sqrt{2} \mathbf{E}(\mathbf{r})  \cos({\omega t})$ of respective RMS intensities $\phi(\mathbf{r}) = \sqrt{\langle \varphi^2(\mathbf{r},t) \rangle}$ and $E(\mathbf{r}) =\sqrt{\langle \mathbfcal{E}(\mathbf{r},t) \cdot \mathbfcal{E}(\mathbf{r},t) \rangle} = ||\mathbf{E}(\mathbf{r})||$, where $\langle\dots\rangle$ denotes a time-average over a period $2\pi/\omega$ and $\mathbf{E}= -\nab \phi$. Uncharged polarizable particles experience a dielectrophoretic force $\mathbf{F_d}$ whose time-averaged expression in the dipole approximation and assuming a spherical shape is:
\begin{equation}
    \langle \mathbf{F_d} \rangle = 2 \pi  R_p^3 \epsilon_m  \Re (K^*)  \boldsymbol{\upxi}
\end{equation}
where $R_p$ is the particle radius, $\epsilon_m$ the medium permittivity, $\Re(K^*)$ the real part of the Clausius-Mossotti factor and $\boldsymbol{\upxi} = \nab E^2$ the gradient of the Mean-Square (MS) electric field $E^2$, that we call "dielectrophretic field" in the following. The  (complex) Clausius-Mossotti factor $K^*$ is defined by: 
\begin{equation}
    K^* = \dfrac{\epsilon_p^*-\epsilon_m^*}{\epsilon_p^*+2 \ \epsilon_m^*} 
\end{equation}
where $\epsilon_p^* = \epsilon_p - i \ \dfrac{\sigma_p}{\omega} $ and $\epsilon_m^* = \epsilon_m - i \ \dfrac{\sigma_m}{\omega}$ are the complex permittivities of the particle and of the medium, respectively. In the case of polystyrene particles in water considered below for numerical applications, the permittivities are $\epsilon_p = 2.55 \ \epsilon_0 $ and $\epsilon_m = 78 \ \epsilon_0 $ with $\epsilon_0 = 8.85 \times 10^{-12}$~F.m$^{-1}$ the vacuum permittivity, while the medium conductivity is $\sigma_m = 0.02$~mS.m$^{-1}$. The conductivity $\sigma_p = \sigma_{p-bulk} + 2K_s/R_p$ of PS particles depends on their radius $R_p$ because it includes contribution from the bulk $\sigma_{p-bulk} = 6.7 \times 10^{-14}$~mS.m$^{-1}$ and surface $2 K_s/R_p$ conductivities, where $K_s = 3.22$~nS is the surface conductance \cite{challier2021printed}. Figure~\ref{fig_geom}(d) summarises the frequency and particle size dependence of the real part of the Clausius-Mossotti factor $\Re(K^*)$. It shows in particular that PS particles with a diameter larger than $5$~µm cannot experience positive DEP (pDEP) in the considered aqueous medium. In this mode, particles are driven toward high electric field regions, \textit{i.e.} at the electrode edges, and their trapping location depends on their initial altitude in the channel (along the $(Oz)$ axis Figure~\ref{fig_geom}(a,b,c)) which prevents the user from controlling the separation condition. Moreover, the electrode edges are high temperature zones due to Joule heating, which can damage/kill living cells. For these two reasons, the pDEP mode is not investigated further and we only consider the negative DEP (nDEP) regime in which particles are driven toward low electric field regions.

Particles also experience a hydrodynamic viscous drag force: 
\begin{equation}
    \mathbf{F_h} = 6  \pi R_p  \mu  (\mathbf{u} - \mathbf{v_p})
\end{equation}
where $\mu$ is the fluid viscosity, $\mathbf{u}$ its velocity field and $\mathbf{v_p}$ the particle velocity. Particle concentration is low such that we neglect particle-particle interactions. The PS particles density is close to the carrier fluid one, so that their sedimentation is ignored (see supplementary section S$2$.$2$ for further discussion). The PS beads we use are larger than $5$~µm, and we neglect their diffusion with respect to advection by the flow and the effect of the dielectrophoretic force. Under these assumptions, from the electrode geometry $(w_e,d_e,\alpha,H)$, voltage $V$, frequency $f$, fluid flow rate $Q$ and the channel geometry $(H,W,L)$, one can predict the trajectories of particles from the forces $\langle \mathbf{F_d} \rangle$ and $\mathbf{F_h} $, and to analyze the possibility of their trapping. To simplify the discussion, we define the reduced RMS potential $\psi$, (opposite) RMS potential gradient $\Tilde{\mathbf{E}}$ and dielectrophoretic $\Tilde{\boldsymbol{\upxi}}$ fields by taking the channel height $H$ as reference length and the RMS electrode voltage $V$ as reference potential: $\psi = \dfrac{\phi}{V}$, $\Tilde{\mathbf{E}}= -H \nab \psi = \dfrac{H}{V} \mathbf{E} $ and $\Tilde{\boldsymbol{\upxi}}= H \nab \Tilde{E}^2 = \dfrac{H^3}{V^2} \boldsymbol{\upxi}$.

\subsection{Numerical modelling}
We apply AC voltages in the low frequency range ($f \leq 10 $ MHz). The associated electromagnetic wavelength is four order of magnitudes larger than the electrode array length. Therefore we ignore the time variations of magnetic induction field $\dfrac{\partial \mathbfcal{B}}{\partial t} \approx \mathbf{0} $ (electroquasistatic framework). As a consequence, Maxwell-Faraday's equation in the medium implies a non-rotational electric field ($\nab \times \mathbfcal{E} = \mathbf{0}$) deriving from the electric potential field $\mathbfcal{E}= - \nab(\sqrt{2} \phi \cos(\omega t)) = \sqrt{2} \mathbf{E} \cos(\omega t) $. The particle concentration being low, we consider the particle suspension as an homogeneous medium of uniform permittivity $\epsilon_m$, conductivity $\sigma_m$ and not carrying charge distribution. Thus the charge conservation equation in the medium and written in the frequency domain  ($\nab \cdot (\sigma_m^* \mathbfcal{E}^*) = 0 $ where $\mathbfcal{E}^* = \sqrt{2} \mathbf{E} \exp(i \omega t)$) reduces to the Laplace equation of RMS potential field $\phi$:
\begin{equation}
    \nabla^2 \phi =0 
\label{eq_lap}
\end{equation}
We neglect the electrode thickness and prescribe $\phi=V$ on the positively charged electrodes (in green and yellow in Figure~\ref{fig_geom}(a,b,c)) and  $\phi=0$ on the grounded electrodes (in red and blue in Figure~\ref{fig_geom}(a,b,c)). On the uncharged boundaries of the channel wall (in grey in Figure~\ref{fig_geom}(b,c)), we consider a perfect insulator boundary condition: $\dfrac{\partial \phi}{\partial n} = -\mathbf{E}\cdot\mathbf{n} = 0 $ where $\mathbf{n}$ is the outward unitary normal vector to the frontier of $\Omega_p$ called $\partial \Omega_p $. Finally, the continuity of $\phi$ is ensured by periodic boundary conditions between the lateral opposite faces of $\partial \Omega_p$ in the flow and flow-transverse directions respectively.  The boundary problem defined by equation~\ref{eq_lap} and the previous boundary conditions are translated into the following weak (integrated) form, that we implement into COMSOL \textregistered \  (see supplementary section S$1$.$2$): 
\begin{equation}
\label{eq_weak}
    \int_{\Omega_p} - \nab \phi  \cdot \nab \hat{\phi} \ d\Omega_p + \oint_{\partial \Omega_p} \dfrac{\partial \phi}{\partial n}  \hat{\phi} \ d \partial \Omega_p = 0
\end{equation}
where $\hat{\phi}$ is any virtual potential field (test function). The first and second integral represent the virtual electrodynamic energy in $\Omega_p$ and its exchange with the surrounding domains of $\Omega_p$ through $\partial \Omega_p$, respectively. Eq.~\ref{eq_weak} is discretised with cubic Lagrange interpolation functions. The resulting linear system is numerically inverted using the COMSOL \textregistered \ direct Multifrontal Massively Parallel Sparse solver (MuMPS).

The microfluidic channel is horizontally orientated such that gravity only induce a vertical pressure gradient not contributing to the flow, thus we neglect the fluid weight. The fluid flow is incompressible, and the low particle concentration and Reynolds number allow to neglect inertial terms of the Navier-Stokes equation. Therefore, in the stationary regime, the velocity $\mathbf{u}$ and pressure $p$ fields satisfy the incompressible Stokes equation:
\begin{equation}
\begin{cases}
    - \nab p+ \mu \nabla^2 \mathbf{u} = \mathbf{0}\\
    \nab\cdot\mathbf{u} = 0
\end{cases}
\label{eq_stokes}
\end{equation}
where $\mu$ is the fluid dynamic viscosity. On the electrode and channel walls, the viscous fluid/structure interaction imposes null velocity: $\mathbf{u} = \mathbf{0}$. We consider the fluid flow as normal to the inlet and outlet boundaries and we neglect the electrode thickness, thus the microfluidic channel has a rectangular cross section and the velocity field solution of equation \ref{eq_stokes} with its boundary conditions is $ \mathbf{u} = u  \mathbf{e_x}$ where $u(y,z)$ is the $(Ox)$ component of fluid velocity field and $\mathbf{e_x}$ is the unitary vector along the $(Ox)$ axis. While $u(y,z)$ can be expressed as a series, since in the present case the microfluidic channel's cross section has a small aspect ratio $\dfrac{H}{W} \sim 0.1$, the fluid velocity field can be well approximated by: 
\begin{equation}
\label{eq_sol_stokes}
    u(z') = U \ \left[1- 4 \left(\dfrac{z'}{H} \right)^2\right]
\end{equation}
where $z'= z-H/2$, $U$ is the maximal fluid velocity located in the $z=H/2$ plane. We neglect the diffusion of colloidal particles with respect to both advection and their dielectrophoretic velocity, so that their dynamics is described by: 
\begin{equation}
\label{eq_traj}
    \dfrac{\partial \mathbf{r}}{\partial t} = \mathbf{u} + m_d  \boldsymbol{\upxi}
\end{equation}
where $\mathbf{r}$ is the particle position and $m_d$ is the particle dielectrophoretic mobility given by: 
\begin{equation}
    m_d=\dfrac{R^2_p \epsilon_m \Re(K^*)}{3 \mu}
\end{equation}
The particle trajectories are computed by integrating equation \ref{eq_traj} using the explicit Runge-Kutta numerical scheme. 

\subsection{Experiments}
High precision capillary printing (Hummink, NAZCA printer) is based on technology borrowed from atomic force microscopy \cite{Guiton_2024}. A micropipette containing the ink to be printed is placed at the end of a macroresonator. A meniscus is formed by approaching the pipette to the surface. When the plate is moved in x and y, the pipette leaves a trail of ink. The printing parameters (movement speed, size of the micropipette), the rheological characteristics of the ink and the ink/substrate adhesion control the width of the printed lines.

Silver electrode arrays were printed on a microscope glass slide with High Precision Capillary Printing (Nazca printer, Hummink) using $5$~µm borosilicate glass micropipettes. The silver nanoparticle ink and the micropipettes were provided by Hummink. The final printed pattern presents interdigitated electrodes ($5$~mm in length, $40$~µm in width and $50$~µm as electrode spacing) covering an area of $5$~mm $\times$ $5$~mm. The printing conditions included a spreading factor of $1.3$, a spiral filling option, and a travel speed of $5000$~µm.s$^{-1}$. Before printing, glass slides were washed with hand soap, then sonicated for $5$~min in acetone. The ink was cured in a convection oven (Memmert) at $180^\circ$C for $1$~hour in air atmosphere. The electrode dimensions were measured using a stylus profilometer (DektakXT, BRUKER), revealing a width of $40$~µm and a height of around $200$~µm. For the final device, the microfluidic channel ($30$~mm$ \times$ $4$~mm, drawn on Autodesk fusion and cut with a xurograph GRAPHTEC CE6000-$40$ plus) was cut out from $49$~µm high double-sided tape (Adhesives Research) which was then placed between two glass substrates with the printed electrode array. To precisely superimpose the two electrodes at $2 \alpha = 90^\circ$, an optical homemade aligner set up was used. The microfluidics connections are made by sticking a $3$D-printed eyelet (Clear resin, FromLab) with a double-sidded tape (DX2, Adhesive Research). More information on the  High Precision Capillary Printing process is provided in supplementary section S$1$.$2$.

In order to validate the simulation predictions, we conducted experiments in distilled water, using PS particles with diameters of $5$~µm and $25$~µm (Duke Standards™ Series $2000$, ThermoFisher, USA). A syringe pump (ISPLab04, DK Infustetek, China) delivers the particle suspension at a flow rate $Q$. An AC voltage $V$ is applied between the electrode arrays in a microfluidic channel using an arbitrary waveform generator (DG$1022$, RIGOL, China) with a peak-to-peak amplitude of $20$~V$_{\text{pp}}$ (\textit{i.e} $V = 7.07$~V in RMS value) and a frequency $f$ of $10$~MHz. This frequency was chosen to ensure an nDEP behaviour of all particles considered in the present work, with a large (absolute) Clausius-Mossotti factor is the largest to maximize the DEP force. In this regime, the CM factor is almost independent of the size, so that the DEP force experience by the particles is proportional to their volume. In addition, this frequency can be easily produced by standard wave generators for a voltage amplitude of $20$~V$_{\text{pp}}$. Since the CM factor saturates at high frequency, it is not necessary to go beyond with more elaborate equipment.

Images and videos are captured with an inverted trinocular microscope (Primovert, ZEISS, Germany) equipped with an MKU Series Color Ocular Camera (The Imaging Source, Germany).

\section{Results and discussion}
\label{sec:ResultsDiscussion}

\subsection{An orthorhombic virtual pillar array}

\begin{figure}[ht!]
    \centering
    \includegraphics[width = \textwidth]{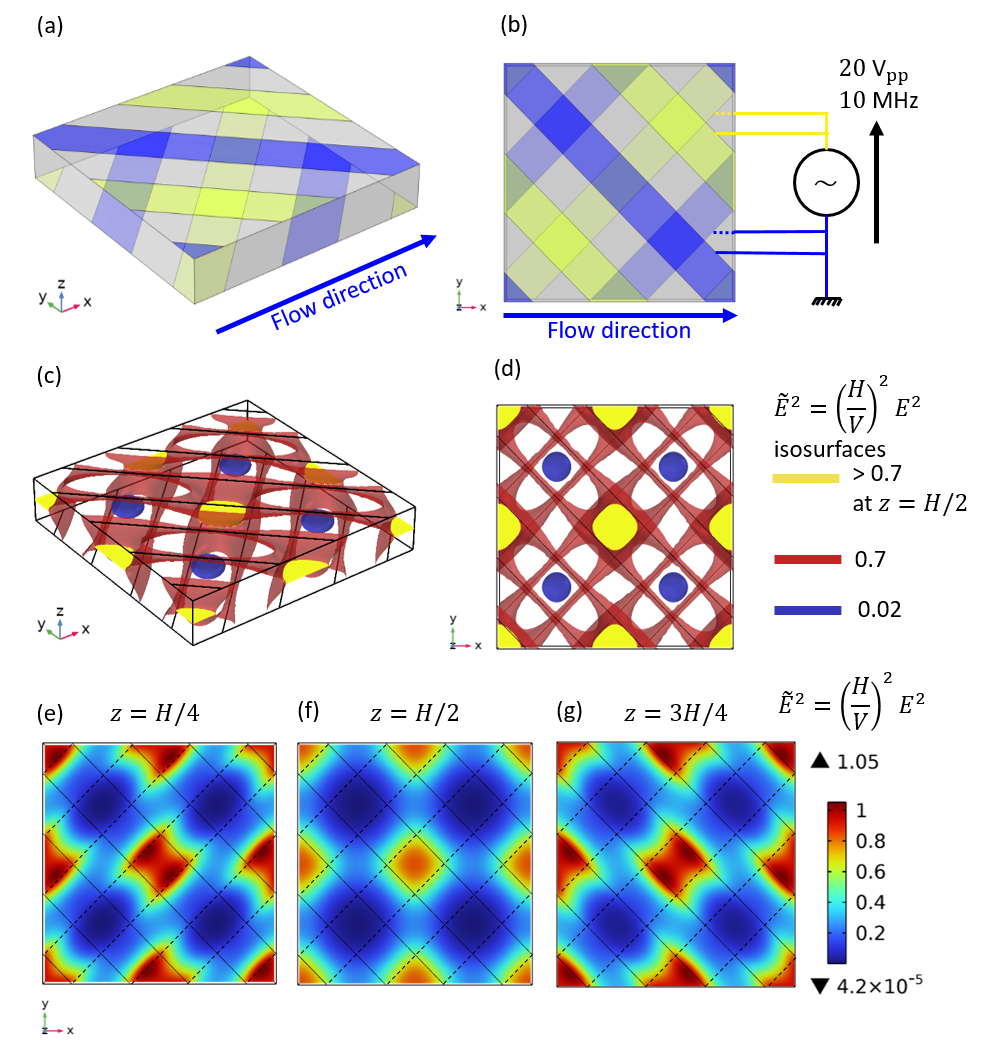}
    \caption{Electrode polarity in the unit cell, shown in perspective (a) and from the top of the device (b); the blue and light green colors indicate electrodes at the same voltage, while grey surfaces on the top and bottom walls indicate insulating regions. Isosurfaces of the reduced MS electric field $\Tilde{E}^2$ for high (red) and low (blue) values viewed in perspective (c) and from the top (d). Color map of the $\Tilde{E}^2$ field at $z=H/4$ (e), $z=H/2$ (central plane, f) and $z=3H/4$ (g). Solid (resp. dashed) black lines indicate the top (resp. bottom) electrode edges.}
    \label{fig_E2}
\end{figure}

In this section, we focus on the electric field generated by the crossed interdigitated electrodes with geometric parameters fixed to respectively: $w_e = d_e = H = 100$~µm and $\alpha = 45 \degree$. The electrode polarity is chosen such that each interdigitated electrode array (top and bottom) has two electrodes of opposite polarities, as illustrated in Figures~\ref{fig_E2}(a) and~\ref{fig_E2}(b), where a voltage $V$ is applied between yellow and blue electrodes. Calculations are performed for $V = 20$~V$_{pp}$, but the reduced MS electric field $\Tilde{E}^2$ does not depend on voltage. Its topology is illustrated in Figures~\ref{fig_E2}(c) and~\ref{fig_E2}(d). The red (resp. blue) surfaces enclose regions where the electric field is large (resp. small), defined by an arbitrary threshold $\tilde{E}^2 >0.7$ (resp. $\tilde{E}^2 <0.02$) for visualization. The high-field regions are also materialized by yellow surfaces indicating their intersection with the channel mid-plane (defined by $z=H/2$). Figure~S1 of the Supplementary Information further illustrates these high-field regions, which extend continuously from the bottom to the top of the channel while remaining  localized in the directions along the planes containing the electrodes. They are approximately perpendicular to both the top and bottom walls and to the direction of the flow. Since they correspond to repulsive zones in nDEP, in the following we will refer to them as ``virtual pillars'' by analogy with steric repulsion by solid obstacles (hence with membrane-based filtration).

The high-field regions are found where electrodes of opposite polarities on each side of the channel "intersect" (see the top views in Figures~\ref{fig_E2}(b) and ~\ref{fig_E2}(d)), \textit{i.e.} where the distance between them is shortest -- and equal to the channel height. As discussed below, these high-field regions form vertical "pillars", which are repulsive for particles in nDEP regime. Their cross-section (in planes parallel to the walls) is approximately parallelepipedic and is determined by the "intersection" of the corresponding electrodes, \textit{i.e.} their width $w_e$ and the angle $2\alpha$ between them, as well as the vertical position $z$ within the channel: the pillars are narrowest in the mid-channel plane and widen closer to the walls, as can be seen in Figures~\ref{fig_E2}(e), \ref{fig_E2}(f) and~\ref{fig_E2}(g). Panels (f) and (g) also show that the magnitude of the field is not uniform within the "pillars": it is smaller in the central plane, where the field is almost in the $z$ direction, and larger close the walls, where the field has a large component in the $xy$-plane because of the voltage between the alternating electrodes on the same wall. This in-plane component is also perpendicular to the electrodes, resulting in maxima of the field near the edges of the electrodes.

In contrast, low-field regions are found where electrodes with the same polarities on each side of the channel "intersect" and are localized around the channel mid-plane (see the top views in Figures~\ref{fig_E2}(b) and ~\ref{fig_E2}(d)). As discussed in more detail below, these ellipsoidal low-field regions act as attractors for particles in the nDEP regime, and constitute stable equilibrium positions for such particles in the absence of fluid flow.

\subsection{Contactless particle capture}

\begin{figure}[ht!]
    \centering
    \includegraphics[]{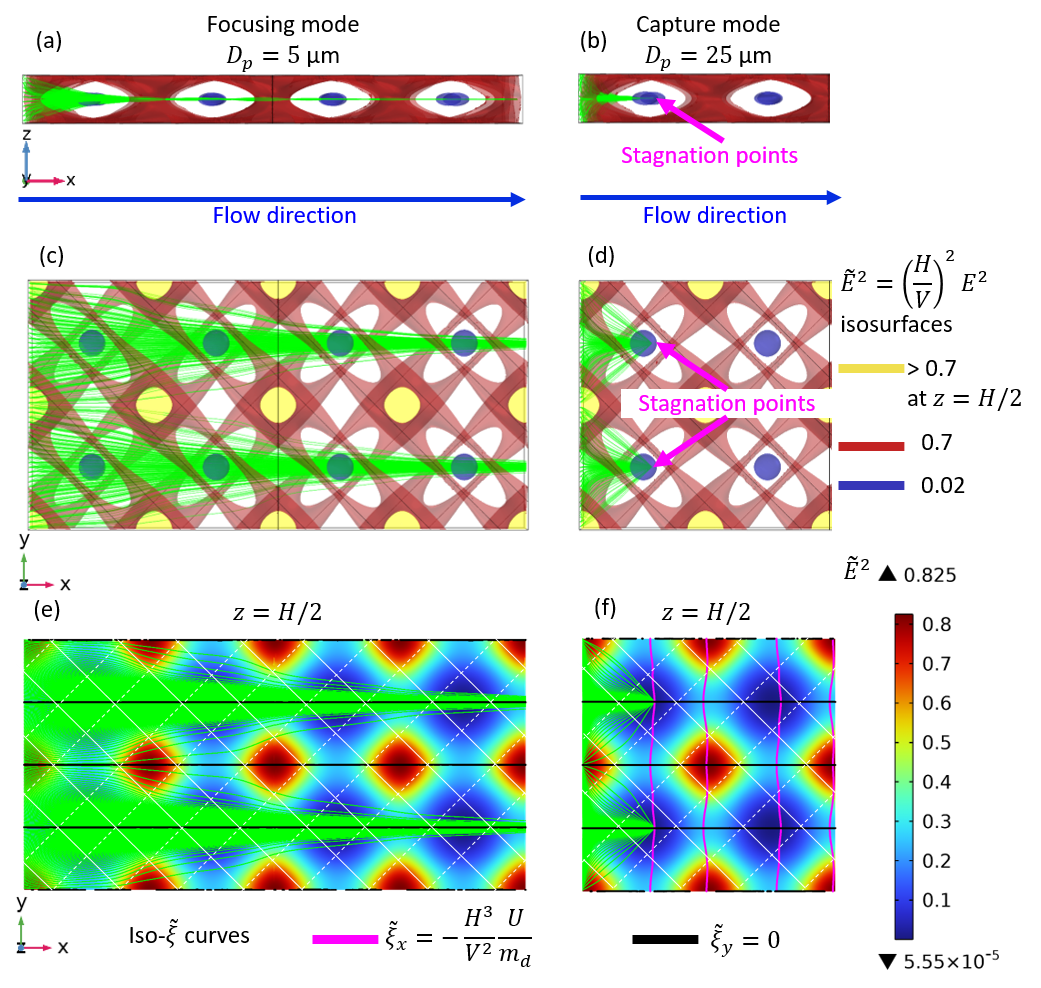}
    \caption{Simulated trajectories of independent polystyrene (PS) beads with diameters $D_p=5$~µm (panels a, c, e) and  $D_p=25$~µm (panels b, d, f) in negative DEP (nDEP) mode for a frequency $f= 10 $~MHz and flow rate $Q = 5$~µl.min$^{-1}$ with $W=4$~mm, $H=50~$µm, $w_e = 40$~µm, $d_e=50$~µm, $\alpha=45\degree$. The trajectories are shown from the side of the channel (panels a and b) and from the top (panels c, d) together with the same isosurfaces of the reduced MS electric field $\Tilde{E}^2$ as in Figure~\ref{fig_E2}. Panels (e) and (f) show the trajectories in the channel mid-plane ($z=H/2)$, together with two isosurfaces of the $x$ (magenta) and $y$ (black) components of the reduced field gradient $\Tilde{\boldsymbol{\upxi}}= H \nab \Tilde{E}^2$. Solid (resp. dashed) white lines indicate the top (resp. bottom) electrode edges in Panels (e) and (f).
    }
    \label{fig_ps_trap}
\end{figure}

In order to illustrate the behaviour of particles in the configuration described previously, we simulate the trajectories of 1000 independent PS beads, for $V = 20$~V$_{pp}$, $f=10$~MHz and a flow rate $Q=5$~µL.min$^{-1}$ in the whole device for $H=50$~µm and $W = 4$~mm, which corresponds here to a maximal flow velocity $U=\dfrac{3}{2} \dfrac{Q}{WH}\approx 0.62$~mm.s$^{-1}$. Two sets of simulations are performed, for two particle diameters $D_p=5$~µm and $25$~µm, both in the nDEP regime at the considered frequency. In each case, particles are initially located on a regular grid in the channel cross section of the unit cell $\Omega_p$. Figures~\ref{fig_ps_trap}(a) and ~\ref{fig_ps_trap}(b) show that in both cases particles rapidly focus in the channel mid-plane ($z=H/2$). Figures~\ref{fig_ps_trap}(c) and ~\ref{fig_ps_trap}(d) further show that they also align laterally ($y$ direction) within distinct parallel corridors between the $\Tilde{E}^2$ pillars, along ligns connecting the field minima in the direction of the flow. However, while all the bigger particles (panels b and d) are stopped at precise stable stagnation points (trapping points) located in the central plane ($z=H/2$) and along the focusing lines (where $y$ is an odd multiple of $W_p/4$), none of the smaller ones (panels a and c) are trapped.

To investigate the existence of stable stagnation points $S(x_s,y_s,z_s)$, we consider the stagnation condition of particles,
\textit{i.e.} try to find the positions where velocity of particles, given by Eq.~\ref{eq_traj}, vanishes. This leads to $\dfrac{\partial \mathbf{r}}{\partial t} = \mathbf{u} + m_d \ \boldsymbol{\upxi}=\mathbf{0}$. Since the Poiseuille fluid flow is unidirectional ($u_y=u_z=0$), projecting the stagnation condition on the three directions of space leads to $u + m_d \ \xi_x=0$ (streamwise), $\xi_y=0$ (lateral) and $\xi_z=0$ (vertical). These three conditions must be satisfied simultaneously. Geometrically, this corresponds to the intersection of the three isosurfaces defined by $\xi_x = -\dfrac{u}{m_d}$, $\xi_y = 0 $ and $\xi_z = 0 $, which can be determined numerically. The vertical projection equation ($\xi_z = 0 $) implies that stagnation points can only exist in the channel mid-plane \textit{i.e} $z_s = H/2$. The lateral projection equation ($\xi_y = 0 $), represented by the black lines in Figs.~\ref{fig_ps_trap}(e) and~\ref{fig_ps_trap}(f) reduces the candidate points to the horizontal lines either linking the pillar corners in the flow direction or linking the $E^2$ minima in the flow direction (the focusing lines). In the mid-plane, the fluid velocity is maximal ($u = U$), so that the streamwise projection of the stagnation condition becomes $U + m_d \ \xi_x = 0 $; it is represented by the magenta curves in Figs.~\ref{fig_ps_trap}(e) and~\ref{fig_ps_trap}(f). 

For particles with the lower DEP mobility (here, smaller diameter), there are no intersections between the relevant $\xi_x$, $\xi_y$ and $\xi_z$ isocurves (because there are no corresponding iso-$\xi_x$ curve), so that particles cannot be trapped there and are simply focused between the pillars (Fig.~\ref{fig_ps_trap}(e)). For particles with the higher DEP mobility (here, larger diameter), the relevant iso-$\xi_x$ curves exist (in magenta) and two types of stagnation points appear (Fig.~\ref{fig_ps_trap}(f)). The first ones, along iso-$\xi_x$ curves going through the virtual pillars ($\Tilde{E}^2$ maxima) are unstable with respect to small displacements along the flow. The second ones,  along iso-$\xi_x$ curves going through the global minima of $\Tilde{E}^2$ (dark blue regions) are stable. This explains why particle trajectories converge and stop at these points in Figs.~\ref{fig_ps_trap}(b), \ref{fig_ps_trap}(d) and~\ref{fig_ps_trap}(f). These observations illustrate the existence of a particle critical mobility $m_{d}^{c}$ between focusing and stagnation mode, corresponding to:
\begin{equation}
\label{eq:mdc}
    m_{d}^{c} = -\dfrac{U}{\xi_{max}} = -\dfrac{H^3}{V^2} \dfrac{U}{\tilde{\xi}_{max}}
\end{equation}
where $\xi_{max} = \max_x(\xi_x(x,y_s,z_s))$ is the maximum of the streamwise component of the DEP field along the focusing lines. Since the maximal velocity $U=\dfrac{3}{2} \dfrac{Q}{WH}$ of the fluid is easily controlled by the hydraulic device, and $\boldsymbol{\upxi} = \nab E^2 $ is fully determined by $H$, $V$ and the reduced $\tilde{\boldsymbol{\upxi}}$, the knowledge of $\tilde{\xi}_{max}$ as a function of the device geometry allows to predict whether a particle can be trapped or not, as we now discuss.

\subsection{Parametric analysis and optimisation}

As mentioned above, for a given device and flow rate, the critical mobility allowing particles to be trapped is determined by the dimensions $H$ and $W$ of the channel, the flow rate $Q$, RMS voltage $V$ and the geometry of the electrode arrays, specifically via $\tilde{\xi}_{max} = \max_x(\tilde{\xi}_x(x,y_s,z_s))$, the maximum of the streamwise component of the (reduced) DEP field along the focusing lines. Therefore, we perform a comprehensive parametric study of the four geometric parameters $\alpha$, $d_e$, $w_e$, $H$ (see Fig.~\ref{fig_geom}c) on $\boldsymbol{\Tilde{\upxi}}$ along the stagnation lines $(x,y_s,z_s)$, from which we extract $\Tilde{\xi}_{max}$. 

\begin{figure}[ht!]
    \centering
    \includegraphics[width=\textwidth]{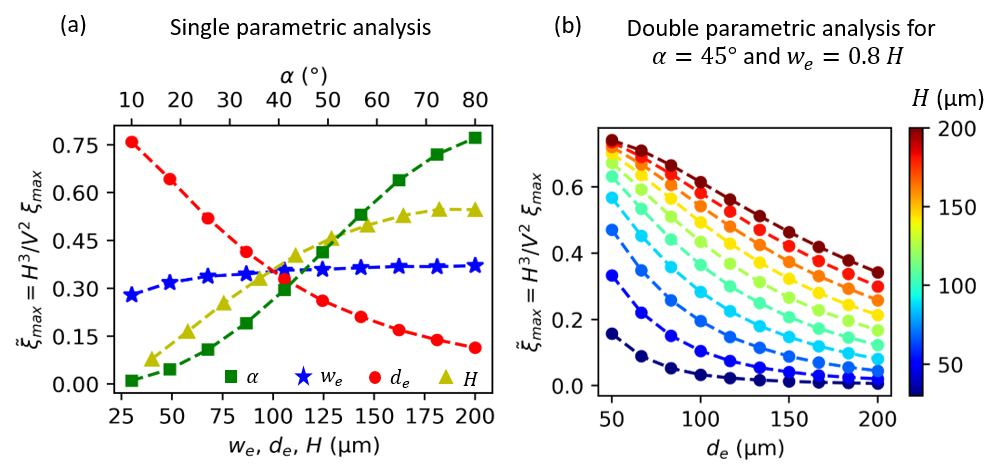}
    \caption{(a) Parametric analysis (see Fig.~\ref{fig_geom}c for the definition of $\alpha, w_e, d_e$ and $H$) of the reduced maximum DEP field along stagnation lines $\Tilde{\xi}_{max}$, varying a single parameter at a time, from the reference configuration $w_e=d_e=H=100$~µm and $\alpha = 45$\degree. Symbols correspond to the numerical results, while dashed lines are guides for the eye. (b) Double parametric analysis of $\Tilde{\xi}_{max}$ when fixing $\alpha=45\degree$ and $w_e= 0.8~H$. Symbols correspond to the numerical results, while dashed lines are guides for the eye. We recall that the actual maximum DEP field is $\xi_{max}=\Tilde{\xi}_{max}V^2/H^3$.
    }
    \label{fig_result_param}
\end{figure}

Figure~\ref{fig_result_param}(a) shows the evolution of $\Tilde{\xi}_{max}$ with each of these parameters, keeping all the others fixed at their reference values ($100$~µm for $w_e, d_e$ and $H$, $45\degree$\ for $\alpha$). In each case, the numerical results (symbol) are shown with lines to guide the eye. We first note that $\Tilde{\xi}_{max}$ only slightly depends on the electrode width $w_e$ (blue stars) in the considered range, in particular for $w_e \geq 100$~µm. When $w_e/H\gg1$, the MS electric field between electrodes of opposite polarities facing each other across the channel is uniform and $E^2 \approx (V/H)^2$. For a fixed electrode spacing, the corresponding gradient $\Tilde{\xi}_{max}$ also converges with $w_e$ (towards approximately $0.4$ in the present case). In fact, this limit is already reached for electrode widths $w_e$ only slightly smaller than the channel height $H$. Since there is no benefit of further increasing $w_e$ in terms of maximum DEP force, and on the contrary this reduces the number of pillars per unit area $n_t = \dfrac{4}{W_p L_p}= \dfrac{\sin(2\alpha)}{2(w_e+d_e)^2}$, we conclude that choosing $w_e = 0.8~H$ is a good compromise.

Figure~\ref{fig_result_param}(a) further shows (red circles) that $\Tilde{\xi}_{max}$ decreases with increasing interelectrode distance $d_e$. This is expected since this also corresponds to increasing the distance between pillars with a fixed $E^2 \approx (V/H)^2$, hence smaller lateral gradients. In addition, $\Tilde{\xi}_{max}$ displays a sinus-like dependence on the electrode crossing angle $\alpha$ (green squares). For $\alpha=45^\circ$, the top and bottom electrode arrays are perpendicular to each other and the trapping sites in the channel mid-plane are square. An angle $\alpha < 45^\circ$ (resp. $\alpha> 45^\circ$) makes the trapping sites and virtual pillars longer in the stream (resp. lateral) direction. When $\alpha \rightarrow 0\degree$, the electrodes align along the flow direction and the resulting electric field gradients are perpendicular to the flow, so that $\xi_x \rightarrow 0$ and $\Tilde{\xi}_{max} \rightarrow 0$. Conversely, when $\alpha \rightarrow 90\degree$, the electrodes align in the lateral direction  and the resulting electric field gradients are parallel to the flow ($\xi_y \rightarrow 0$), which maximizes $\Tilde{\xi}_{max}$. However, a too large $\alpha$ value significantly increases the size of trapping sites in the lateral direction and drastically decreases the trapping site surface density $n_t = \dfrac{\sin(2\alpha)}{2(w_e+d_e)^2}$. Therefore a value $\alpha\approx45\degree$ provides a good compromise for the overall efficiency of the setup, even though other values can be adopted to modulate the properties of the device.

Finally, Figure~\ref{fig_result_param}(a) (yellow triangles) shows the influence of $H$ on $\Tilde{\xi}_{max}$. Since $H$ varies in this parametric study, it is important to note that the reference dielectrophoretic field $V^2/H^3$ also varies. For small $H$, ($40$~µm $\leq H \leq$ $80$~µm), $\Tilde{\xi}_{max}\propto H$, so that $\xi_{max}\propto V^2/H^2$. For larger $H$ ($80$~µm $\leq H \leq$ $200$~µm), $\Tilde{\xi}_{max}$ plateaus near $0.4$, so that $\xi_{max}\propto V^2/H^3$. As a result, among the various parameters, for the chosen reference configuration $H$ is the one that has the largest influence on $\xi_{max}$. It should be as small as possible to maximize the dielectrophoretic effects.

Considering the previous analysis, we suggest to adjust the electrode width $w_e$ to be slightly smaller than the channel height using $w_e=0.8~H$ and to fix the crossing angle $\alpha$ to $45^\circ$. The fabrication process constrains both the electrode spacing $d_e$ and the channel height $H$. The latter also depends on the size of the colloidal particles that need to be transported in the liquid. Figure~\ref{fig_result_param}(b) illustrates how the resulting $\Tilde{\xi}_{max}$ depends on both $d_e$ and $H$. We recall that the actual maximum DEP field is $\xi_{max}=\Tilde{\xi}_{max}V^2/H^3$. These results can be used to optimize the setup to tune the critical mobility.

\subsection{Membrane-less dielectrophoretic microseparation}

\begin{figure}[H]
    \centering
    \includegraphics[]{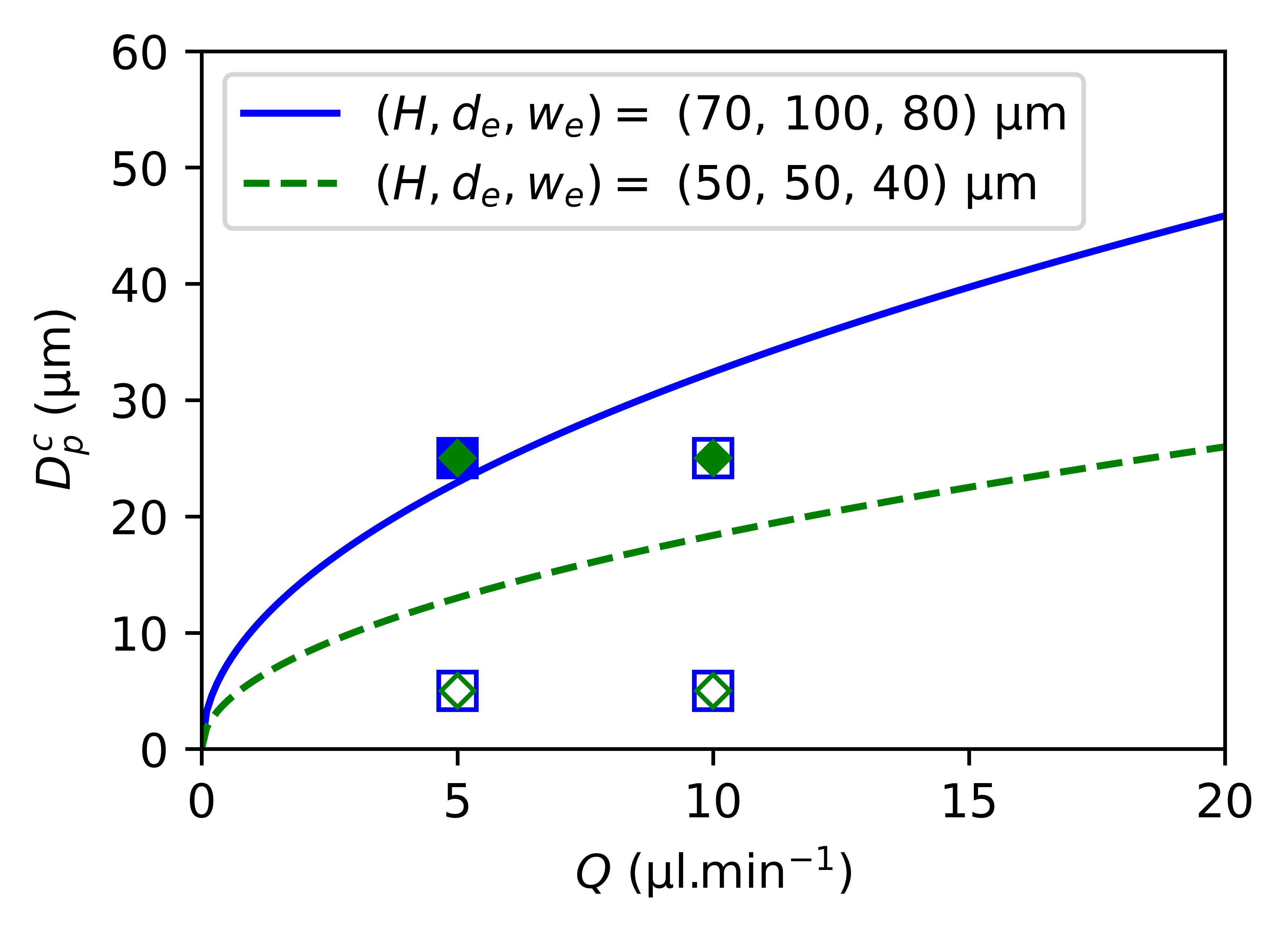}
    \caption{Critical diameter $D^c_p$ of polystyrene particles as function of the flow rate $Q$, for a microfluidic channel of width $W=4$~mm, an angle between the electrodes $2\alpha=90\degree$, and an electrode peak-to-peak voltage of $20$~V$_{\text{pp}}$ (\textit{i.e} $V=7.07$~V). The blue solid (resp. green dashed) line corresponds to a channel height $H=70$~µm (resp. $50$~µm), an electrode spacing $d_e=100$~µm (resp. $50$~µm) and an electrode width $w_e=80$~µm (resp. $40$~µm). The symbols indicate the predicted behaviour of particles with diameters of $5$ and $25$~µm with flow rates $Q =5$ and $10$~µl.min$^{-1}$: full (resp. open) symbols for trapped (resp. not trapped) particles with the first (squares) and second (diamonds) setup. 
    }
    \label{fig_Dpc}
\end{figure}

We can now illustrate the principle of membrane-less dielectrophoretic microseparation using the proposed setup. The critical mobility allowing particles to be trapped by the virtual pillar array, Eq.~\ref{eq:mdc}, corresponds to a critical particle diameter $D_{p}^c$ such that $m_{d}^{c} =(D_{p}^c)^2 \epsilon_m \Re(K^*)/12 \mu$.  In a Hele-Shaw cell, the fluid velocity $U$ in the channel mid-plane (which is also the maximal velocity) can be expressed from the fluid flow rate, the channel width and its height as $U = \dfrac{3}{2} \dfrac{Q}{W H}$. Therefore, the critical particle diameter for trapping by nDEP is:
\begin{equation}
\label{eq:dpc}
    D_{p}^c = \sqrt{\dfrac{18 \mu }{\epsilon_m |\Re(K^*)|}} \times \sqrt{\dfrac{Q}{W \Tilde{\xi}_{max}} } \times \dfrac{H}{V}
    \; ,
\end{equation}
where the first term depends on the voltage frequency, particle and medium properties, whilst the second and third terms include only controlled device parameters, namely geometry (see Figure \ref{fig_result_param}(b)), flow conditions and voltage intensity. For particles larger than $5$~µm and for a frequency $f=10$~MHz, the first term does not depend on the particle size (see Figure~\ref{fig_geom}(d)) and the critical diameter can be represented as a function of the device flow rate $Q$ for a fixed device geometry, as illustrated in Figure~\ref{fig_Dpc}. For example (blue solid line), we predict that using a non-optimised geometry with $W=4$~mm, $H=70~$µm, $w_e = 80$~µm, $d_e=100$~µm, and $\alpha=45\degree$, with an electrode RMS voltage $V = 7.07$~V (corresponding to $20$~V$_{pp}$), $25$~µm particles will not be captured at a flow rate $Q = 10$~µl.min$^{-1}$. They would nevertheless be focalized in the channel mid-plane and along the focusing lines. Such particles should however be trapped at the same flow rate, in a channel with $W=4$~mm, $H=50~$µm, $w_e = 40$~µm, $d_e=50$~µm, $\alpha=45\degree$ (green dashed line), unlike smaller $5$~µm particles. Another option to trap the $25$~µm particles, using the first setup, is to reduce the flow rate to $Q = 5$~µl.min$^{-1}$. The predicted behaviour of particles with diameters of $5$ and $25$~µm with flow rates $Q =5$ and $10$~µl.min$^{-1}$ for the two setups is indicated by symbols in Figure~\ref{fig_Dpc}.

\subsection{Experimental validation}
\label{sec:experiments}

\begin{figure}[ht!]
    \centering
    \includegraphics[]{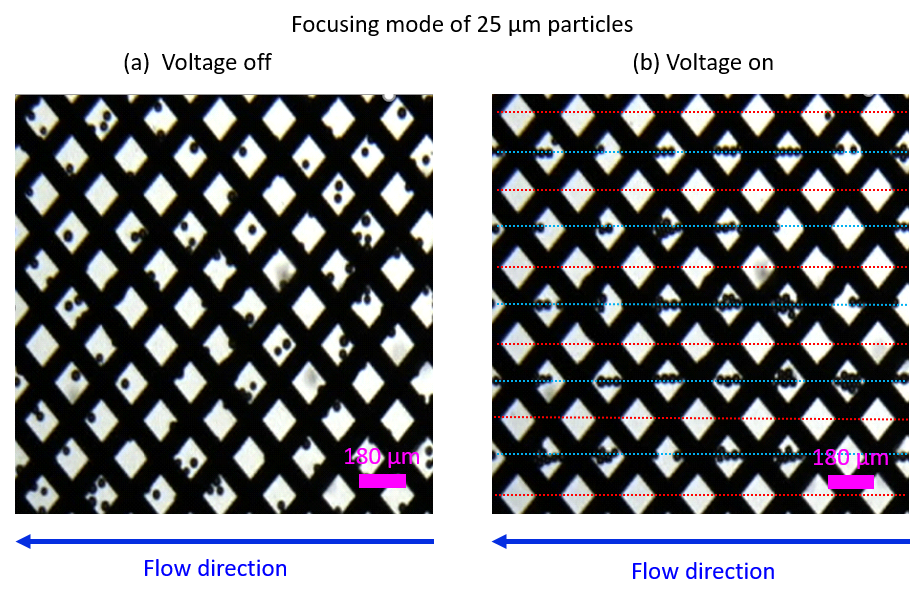}
    \caption{Observation of the focusing mode for $25$~µm diameter PS beads at a voltage frequency $f=10$~MHz and flow rate $Q= 10$~µl.min$^{-1}$ in a non-optimised device with microfluidic channel height $H=70$~µm and width $W=4$~mm, electrode width and spacing are $w_e=80$~µm and $d_e=100$~µm, respectively, and angle between electrodes $2\alpha = 90\degree$ (corresponding to the solid blue line in Figure~\ref{fig_Dpc}). (a) Particle distribution in the device without electrode voltage ($V=0$~V). (b) Particle distribution in the device after applying a voltage of $20$~V$_{\text{pp}}$. Red dotted lines indicate the location of "virtual pillars" of high $E^2$ field along the stream (see Figure~\ref{fig_ps_trap}), whilst blue dotted lines indicate particle focusing lines connecting $E^2$ field minima along the stream (see Figure~\ref{fig_ps_trap}).}
    \label{Fig_exp_focus}
\end{figure}

Figure~\ref{Fig_exp_focus} displays the distribution of $25$~µm particles within the device for a non-optimised geometry with the following parameters: $w_e=80$~µm, $d_e=100$~µm, $\alpha=45\degree$, $H=70$~µm and $W=4$~mm. The corresponding movie is provided as supplementary movie~1.  Panel~\ref{Fig_exp_focus}(a), illustrates the particle distribution before applying voltage, which does not exhibit any specific feature. In contrast, Panel~\ref{Fig_exp_focus}(b) shows that when voltage is applied (after $t \approx 2$~s in the supplementary movie $1$), particles focus along horizontal lines along the stream (blue dotted lines). This lateral focusing process takes approximately $3$~s to complete (from $t \approx 2$~s to $t \approx 5$~s in supplementary movie $1$). However, the particles are not stopped within the flow and finally cross the virtual DEP pillar array because the particle diameter is smaller than the calculated critical diameter, namely $D^c_p = 32$~µm in that case (see Figure~\ref{fig_Dpc}). These particle alignments correspond to the focusing mode previously revealed by the numerical simulations (see Figure~\ref{fig_ps_trap}(a,c,e)). In addition, we note the absence of particles between the focusing lines (red dotted lines), consistently with the predicted presence of $E^2$ pillars that repel particles towards the focusing lines. Finally, we note that the velocity of particles along the stream significantly increases after switching on the generator (see supplementary movie $1$ for $t\geq 2.0$~s). This provides an indirect evidence of the vertical focusing in the channel mid plane, where the fluid velocity is maximal ($u=U$), predicted by the simulations (see Figure~\ref{fig_ps_trap}(a)).

\begin{figure}[ht!]
    \centering
    \includegraphics[]{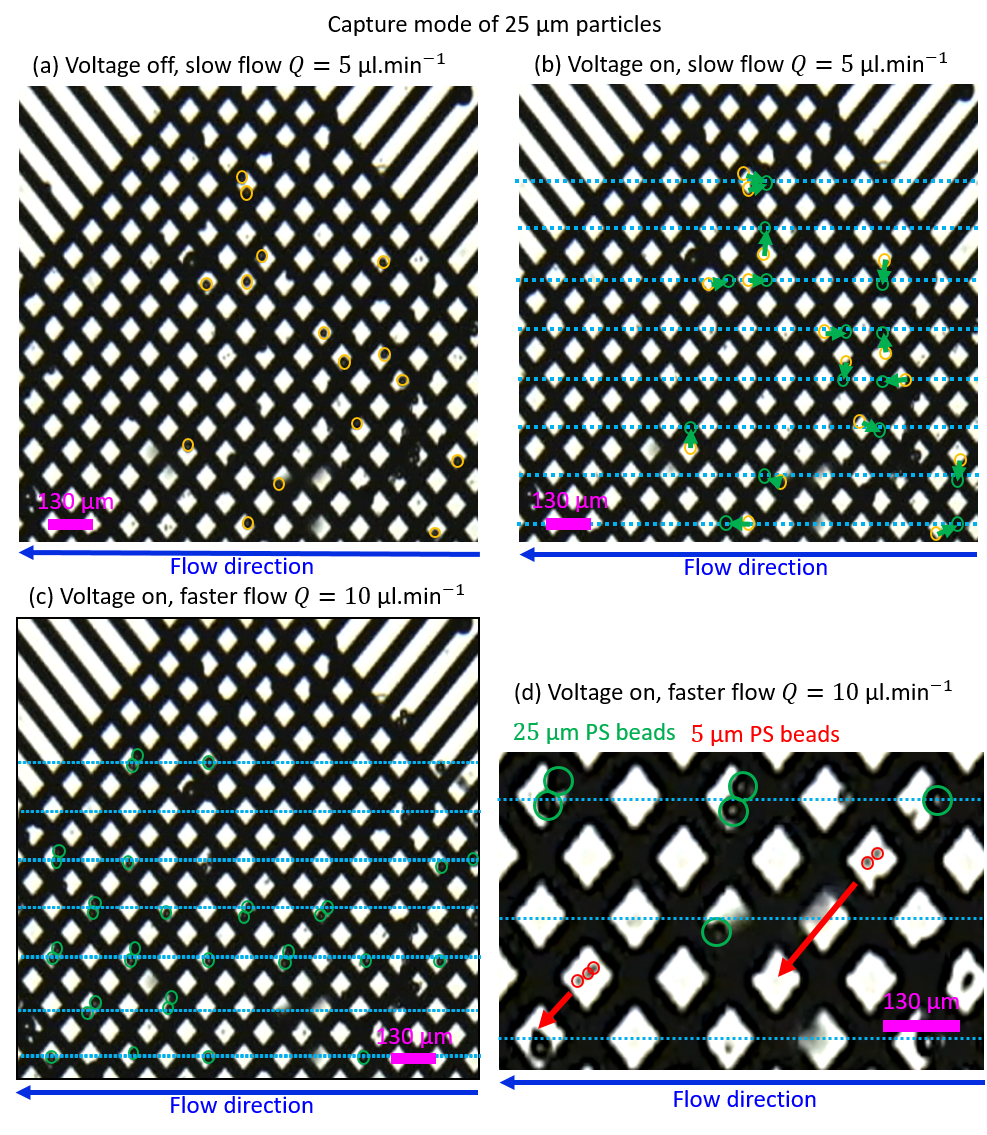}
    \caption{Observation of the capture mode of PS beads at a voltage frequency $f=10$~MHz in an optimised device with microfluidic channel height $H=50$~µm and width $W=4$~mm, electrode width and spacing are $w_e=40$~µm and $d_e=50$~µm, respectively, and angle between electrodes $2\alpha = 90\degree$ (corresponding to the dashed green line in Figure~\ref{fig_Dpc}). (a) Particle distribution before applying voltage at a low flow rate $Q = 5$~µl.min$^{-1}$. Some visible yet untrapped $25$~µm particles are circled in orange. (b) Particle positions when the generator is switched on with $V=20$ V$_{\text{pp}}$. The green circles indicate the location of particles between electrodes (hence not visible from the top), with green arrows from their initial positions, identical to panel (a). (c) Capture of $25$~µm particles with a higher flow rate $Q = 10$~µl.min$^{-1}$. (d) Zoomed view on $5$~µm particles, circled in red, with red arrows indicating the direction of their displacement.}
    \label{fig_exp_capture}
\end{figure}

Figure~\ref{fig_exp_capture} shows the distribution of $25$~µm and $5$~µm particles within the device for an optimised geometry with the following parameters: $w_e=40$~µm, $d_e=50$~µm, $\alpha=45\degree$, $H=50$~µm and $W=4$~mm at different times and flow rates. The associated movie is provided as supplementary material. 
In the absence of voltage (supplementary movie $2$ for $ 0  \leq t \leq 3.66$~s), particles follow the flow and some of them, visible between the electrodes yet untrapped, are shown in panel~\ref{fig_exp_capture}(a) for a low flow rate $Q = 5$~µl.min$^{-1}$. Once voltage ($20$~V$_{\text{pp}}$) is turned on (supplementary movie $2$ for $ 3.66 \leq t \leq 10.00$~s), particles are very rapidly trapped where electrodes "intersect" in top view, along the focusing lines represented by blue dashed lines in panel~\ref{fig_exp_capture}(b). The green circles indicate the location of particles between electrodes (hence not visible from the top), with green arrows from their initial positions, identical to panel~\ref{fig_exp_capture}(a). Such a capture is consistent with the predictions, since for these conditions the calculated critical diameter is $D^c_p = 13$~µm (see green dashed line in Figure~\ref{fig_Dpc}).

\clearpage

After switching off the generator (at $t \approx 10.00$~s in the supplementary movie $2$), the flow rate is increased to $Q = 10$~µl.min$^{-1}$ and the generator is switched back on at the same voltage as previously (at $t\approx12.77$~s). Panel~\ref{fig_exp_capture}(c) shows the particle distribution for this faster flow. As for the smaller flow rate, $25$~µm particles are trapped (sometimes by pair), consistently with the prediction of a critical diameter $D^c_p = 18$~µm for $Q = 10$~µl.min$^{-1}$, still smaller than $25$~µm (see green dashed line in Figure~\ref{fig_Dpc}). However, since the flow rate is increased, the $25$~µm particle equilibrium position is located further downstream with respect to the $E^2$ minima at electrode "intersections" where the streamwise component of the dielectrophoretic field $\xi_x$ is greater and closer to its maximum $\xi_{max}$ along the focusing lines. As a result, the $25$~µm particles are now visible in the spacing between the electrodes. In contrast, smaller particles of $5$~µm are not trapped at this flow rate, as illustrated in the zoom of Panel~\ref{fig_exp_capture}(d), again consistently with the prediction of $D^c_p = 18$~µm. The direction of their motion is indicated by red arrows, which suggest that they experience a force similar to that induced by "virtual deterministic lateral displacement"~\cite{collins2014particle} because they are closer to one of the two electrode arrays. However, as they continue their journey through the device they are eventually focalized in the channel midplane and along the focusing lines.

\subsection{Discussion}
\label{sec:discussion}

Using the relation between the dielectrophoretic barrier and the device parameters  established in this work, the present system could also be used to characterize the dielectrophoretic mobility of particles. A corresponding experimental protocol would begin by trapping particles in the $\Tilde{E}$ minimums (in the centres of blue ellipsoids previous Figure~\ref{fig_E2}(c),(d)). In a second step, the flow rate could be slowly increased until the initially trapped particles cross the $\Tilde{E}$ barrier between two trapping sites at a critical flow rate $Q^c$. The corresponding particle dielectrophoretic mobility is: $m_d = -3H^2Q^c/2WV^2 \Tilde{\xi}_{max}$, where $\Tilde{\xi}_{max}$ is given by Figure~\ref{fig_result_param}(b). For the same purpose, another protocol could be based on slowly decreasing the electrode RMS voltage $V$ while maintaining a constant flow rate until particles cross the $\Tilde{E}$ barrier at a critical RMS voltage $V^c$. In this case, the particle dielectrophoretic mobility is: $m_d = -3H^2Q/2W(V^c)^2 \Tilde{\xi}_{max}$. 

The experimental validation obtained with PS particles opens the way to the use of the proposed model to directly tailor the device characteristics (channel length, thickness and spacing of printed lines) to the real system of interest to be separated. Thanks to the printing methods, these parameters can be adjusted over a wide range to avoid the saturation of the traps created by the virtual pillars. We previously showed that large printed surfaces, with long and paralleled microchannel with dozens of electrode patterns per mm$^2$ can be produced~\cite{challier2021printed}. Furthermore, the proposed setup can be used not only in a continuous-flow process, but also with time-dependent protocols, whereby some particles are selectively trapped, then released (by turning voltage off) to free the traps for the analysis of additional sample of the same solution.

We note that the same trapping condition principle remains the same for more complex particles $m_d^c = -\dfrac{U}{\xi_{max}}$. However, the relation between the critical mobility and the particle geometry and electric properties can change. For a spherical cell, a simple single shell model could be used to account the heterogeneity of the electric properties which modifies the expression of $\Re(K^*)$ where $K^*$ depends on the membrane, cytoplasm and medium properties \cite{ko2004dielectric}. For non-spherical particles such as prolate ellipsoids (with long radius $a$ and short radius $b$): the time averaged DEP force is \cite{DEP_ellipsoide} $\langle\mathbf{F_d}\rangle = 2 \pi a b^2 \epsilon_m \left[\Re(K_x^*) \boldsymbol{\nabla} E_x^2 + \Re(K_{y}^*) \boldsymbol{\nabla}E_y^2 + \Re(K_{z}^*) \boldsymbol{\nabla} E_z^2 \right]$. Therefore, particles experiencing nDEP still focus along the lines where $E_x^2 \approx E^2$ (because of the source symmetry). Moreover, the DEP torque aligns the longest axis of the ellipsoid with the largest field gradient direction \textit{i.e.} along the focusing lines, therefore:  $\langle \mathbf{F_{d}}\rangle \approx 2 \pi a b^2 \epsilon_m \Re(K_x^*) \boldsymbol{\upxi}$. For this forced orientation, the viscous drag force along the focusing lines is $\mathbf{F_h} = 6 \pi a C_d \mu (\mathbf{u}- \mathbf{v_p}) \mathbf{e_x}$, where $C_d$ is the drag coefficient that depends on $a$ and $b$ so that the critical mobility is $m_d^c =\dfrac{b^2 \epsilon_m \Re (K_x^*)}{3 \mu C_d}$.

\section{Conclusion and Perspectives}

We report a new concept of membrane-less separation of micro-particles in microfluidic devices, using two pairs of crossed interdigitated electrodes to control the dielectrophoretic force field throughout the channel. This design avoids the drawbacks of physical membranes such as high hydraulic resistance and the limited electric field frequency range available in the iDEP mode. These two features are particularly relevant for the analysis of large volumes (typically $10$ to $100$~mL) of dilute colloidal suspensions with a variety of particle sizes (diameter $\sim 5-50$~µm) and materials (\textit{e.g.} polymers, biological cells, minerals$\dots$). The device can be easily fabricated on very wide surfaces e.g by electrode High Precision Capillary Printing. Moreover, we describe here the advantage of simple virtual pillars generation across the microchannel section over superimposed and aligned electrode pattern: this new design is easily manufactured, without any need of a complex electrode alignment systems. A parametric analysis showed how the system geometry can be tuned in order to selectively capture particles with given properties in a mixture. This feature was illustrated here on how to separate polystyrene particles with different sizes, but it could also be applied to separate particles with similar sizes and different electric properties or shapes. Importantly, we provided an experimental validation of the proposed ideas, thereby demonstrating the possibility to separate particles according to their size using an array of virtual pillars, induced by the electrode arrays on the top and bottom of the microfluidic channel. The next step of this study is to perform the separation of planktonic cells by size/shape using the same device.

The present theoretical description of the system relies on several assumptions. Firstly, the time averaged expression of the dielectrophoretic force assumes that the particle's diameter is smaller than the scale of non-uniformity of the applied electric field (without particle). Secondly, the resulting particle polarization is considered as a dipole moment and higher order multipoles are neglected. Finally, the Stokes drag force is also implemented considering that the particle is small with respect to scale over which the fluid velocity field varies. All these assumptions could be tested numerically for a given system, but we leave this for further study. This would require in particular to introduce explicit particles in the calculations and, for example, the actual dielectrophoretic net force could be calculated by integrating the Maxwell stress tensor on the particle's surface and the net hydrodynamic force by integrating the fluid Cauchy stress tensor on the same surface. 

The proposed setup can be tailored to target particles with specific properties thanks to the present theoretical analysis, the possibility to fabricate the electrode patterns over large surfaces using High Precision Capillary Printing, with variable geometry, and to control the voltage magnitude and frequency. One can therefore design microfluidic devices including several stages of such membrane-less dielectrophoretic sieves, where different particles will be trapped or simply slowed down according to their dielectrophoretic mobility, while the other particles will pass almost unaffected and separated further downstream with similar stages with different properties. Furthermore, one can combine such spatial design (varying electrode width, $w_e$, and spacing, $d_e$, as well as voltage magnitude, along the microfluidic channel) with temporal sequences of applied voltage in order to selectively trap then release particles (as in steric exclusion chromatography) according to their DEP mobility. This opens the possibility to deal with large volumes of complex samples with a high throughput, without suffering from the usual problems of membrane-based processes such as pressure drop and fouling.

Finally, once isolated, single particles in the DEP traps can be investigated by DEP of other analytical (\textit{e.g.} optical) techniques for sub-population analysis. We are currently developing a dielectrophoretic device dedicated to the characterisation of subpopulations of Alexandrium minutum (a toxic microorganism present in the marine environment) using its dielectrophoretic crossover frequency as it grows. This use case will allow us to assess the precise analytical capabilities in terms of capture rate and selectivity, as well as long-term chemical or mechanical stability, which will be investigated in future work. We note that in our previous work using similar electrode material~\cite{challier2021printed}, with a different design resulting in DEP traps on the top/bottom walls, no electrode dissolution was observed for frequencies larger than 1~kHz (much smaller than the 10~MHz used here), at the same voltage of 20~V$_{\rm pp}$. Therefore we do not expect (and indeed did not observe during the time over which the experiments were performed) such a phenomenon to be a limiting factor in the present case.

\section*{Data availability}
The data corresponding to the figures of this article will be available on Zenodo at \url{https://zenodo.org/records/15103277}. The movies corresponding to the figures and discussed in the text are  available as Supplementary Information.

\section*{Conflict of interest}
There are no conflicts of interest to declare.

\section*{Acknowledgements}
This project received funding from the ANR (grant number ANR-21-CE29-0021-02) and from the European Research Council under the European Union’s Horizon 2020 research and innovation program (grant agreement no. 863473).

\bibliographystyle{rsc}
\bibliography{bib_article.bib}

\section{Supplementary information }

\subsection{High Precision Capillary Printing}
The High Precision Capillary Printing technology introduced by Hummink's Nazca printer involves three elements: a macroresonator (the centimetric tuning fork), a glass nanopipette filled with silver ink (Hummink), and an electronic feedback loop derived from (a phase-locked loop). The macroresonator tuning fork plays the same role as an AFM cantilever with the difference that adding a glass pipette on one prong of the fork does not affect its resonance. The feedback loop allows one to maintain a soft contact between the pipette’s tip and a substrate without damaging the tip \cite{Canale_2018}. The pipette can be seen as a nano-fountain pen printing any kind of ink on any substrate. Upon contact between the pipette and the substrate, an ink meniscus forms. With an appropriate formulation of the ink, the displacement of this meniscus leaves a thin film of material behind. The pipettes are pulled from borosilicate glass capillaries ($1$~mm outer diameter, $0.7$~mm inner diameter, $10$~mm length, WPI) with a P2000 machine (Sutter Instruments).

\subsection{Finite-Element implementation of the electroquasistatics problem}
\label{sec_SI_weak}
We multiply the Laplace equation by a test function $\hat{\phi}$ and integrate it on $\Omega_p$: 
\begin{equation}
    \int_{\Omega_p} \hat{\phi} \nabla^2 \phi \ d\Omega_p =0 \ .
\end{equation}
Then, using $\nab\cdot(\hat{\phi} \nab \phi ) = \hat{\phi}\nabla^2 \phi  + \nab \phi\cdot\nab \hat{\phi}$: 
\begin{equation}
    \int_{\Omega_p} - \nab \phi \cdot \nab \hat{\phi} \ d\Omega_p
    +\int_{\Omega_p} \nab\cdot( \hat{\phi} \nab \phi ) \ d\Omega_p  
     =0 
\end{equation}
and the divergence theorem simplifies the second integral, leading to the weak form equation that we implement in COMSOL:
\begin{equation}
    \int_{\Omega_p} - \nab \phi\cdot\nab \hat{\phi} \  d\Omega_p + \oint_{\partial \Omega_p} \dfrac{\partial \phi}{\partial n} \ \hat{\phi} \ d\partial\Omega_p = 0
    \label{weak_form}
\end{equation}
The first domain integral represents a bilinear, continuous and coercive application of $\phi$ and $\hat{\phi}$. By applying the boundary conditions, the second integral term becomes a linear and continuous application of $\hat{\phi}$. Then, the Lax-Milgram theorem ensures unicity and continuity of the potential field $\phi$ solution of the problem that we discretize by cubic Lagrange interpolation polynomials. By applying this (Galerkin) approximation of $\phi$ and replacing the test function by all interpolation functions, Eq.~(\ref{weak_form}) becomes a linear system that we solve numerically with the COMSOL MuMPS solver.

\subsection{Parameter values}

\begin{table}[ht]
\begin{footnotesize}
     \begin{center}
     {\setlength\arrayrulewidth{1pt}     \begin{tabular}{c c c}
      \hline
      Geometry before optimisation & Description & Value\\
      \hline
      $w_e$ & Electrode width & $100$~µm \\
      $d_e$ & Electrode spacing & $100$~µm\\  
      $\alpha$ & Electrode-flow angle & $45$°\\
      $W$ & Channel width & $3$~mm\\ 
      $H$ & Channel height & $ 100$~µm\\ 
      \hline
      Microfluidic/electric device & Description & Value \\
      \hline
      $f$ & Voltage frequency & $10$~MHz \\
      $\mu $ & Fluid dynamic viscosity & $10^{-3}$~Pa.s \\  
      $Q$ & Flow rate & $5$-$10$~µl.min$^{-1}$\\ 
      $\rho$ & Fluid density & $10^3$~kg.m$^{-3}$ \\ 
      $V$ & Electrode voltage & $20$~V$_{pp}$ ($7.07$~V RMS) \\
      \hline
      Microparticle properties & Description & Value \\
      \hline
      $\epsilon_{m}$ & Medium permittivity & $ 78 \ \epsilon_0$  \\ 
      $\epsilon_{p}$ & PS particle permittivity & $ 2.5 \ \epsilon_0$  \\ 
      $\sigma_{m}$ & Medium conductivity & $0,02$~mS.m$^{-1}$  \\ 
      $\sigma_{p-bulk}$ & PS particles bulk conductivity & $6.7 \times 10^{-14}$~mS.m$^{-1}$ \\ 
      $K_s$ & PS particles surface conductance & $3.22$~nS \\
      \hline
      \end{tabular}
      }
      \caption{Model parameters used in the reference configuration. }
      \label{tableau_all_param}
      \end{center}
      \end{footnotesize}
\end{table}

\subsection{Zoom on a nDEP ``virtual pillar''}

\begin{figure}[H]
    \centering
    \includegraphics{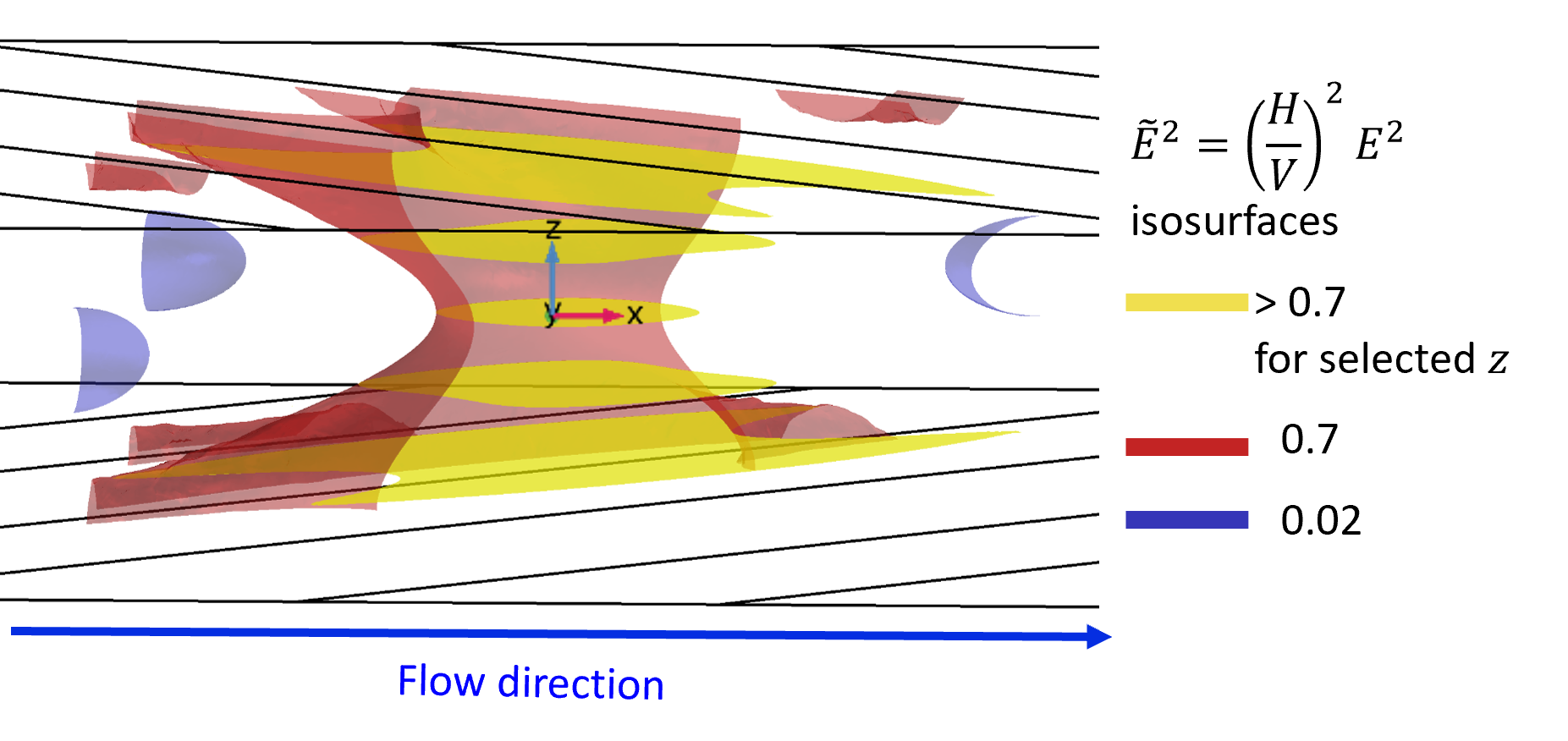}
    \caption{Zoom on a single nDEP ``virtual pillar'' for $(x,y) \in [L_p/4, 3L_p/4] \times [W_p/4, 3W_p/4]$. The red isosurface is cut along the bottom electrode direction to facilitate visualization. Intersections of the $\tilde{E}^2 >0.7$ volume with selected planes parallel to the top and bottom channel walls ($\{z = kH/6 \quad k \in  \llbracket 1, 5 \rrbracket \}$) are shown as yellow surfaces.}
    \label{fig_zoom_pillar}
\end{figure}

As described in Section~3.1 of the main text, high electric field regions within the device act as obstacles to particles experiencing nDEP. Figure~\ref{fig_zoom_pillar} further illustrate the shape of these regions shown \textit{e.g.} in Figures~2(c) and~2(d) of the main text, focusing on a single of them. The red (resp. blue) surfaces enclose regions where the electric field is large (resp. small), defined by an arbitrary threshold $\tilde{E}^2 >0.7$ (resp. $\tilde{E}^2 <0.02$) for visualization. In order to better see the interior of these regions, only a fraction of the red isosurface is shown, and the intersections of the $\tilde{E}^2 >0.7$ volume with selected planes parallel to the top and bottom channel walls (\textit{i.e.} fixed $z$) are shown as yellow surfaces. This representation highlights the fact that the high-field regions extend continuously from the bottom to the top of the channel while remaining  localized in the directions along the planes containing the electrodes. They are approximately perpendicular to both the top and bottom walls and to the direction of the flow. Since they correspond to repulsive zones in nDEP, we refer to them as ``virtual pillars'' by analogy with steric repulsion by solid obstacles (hence with membrane-based filtration).

\subsection{Particle sedimentation}
\begin{figure}[H]
    \centering
    \includegraphics{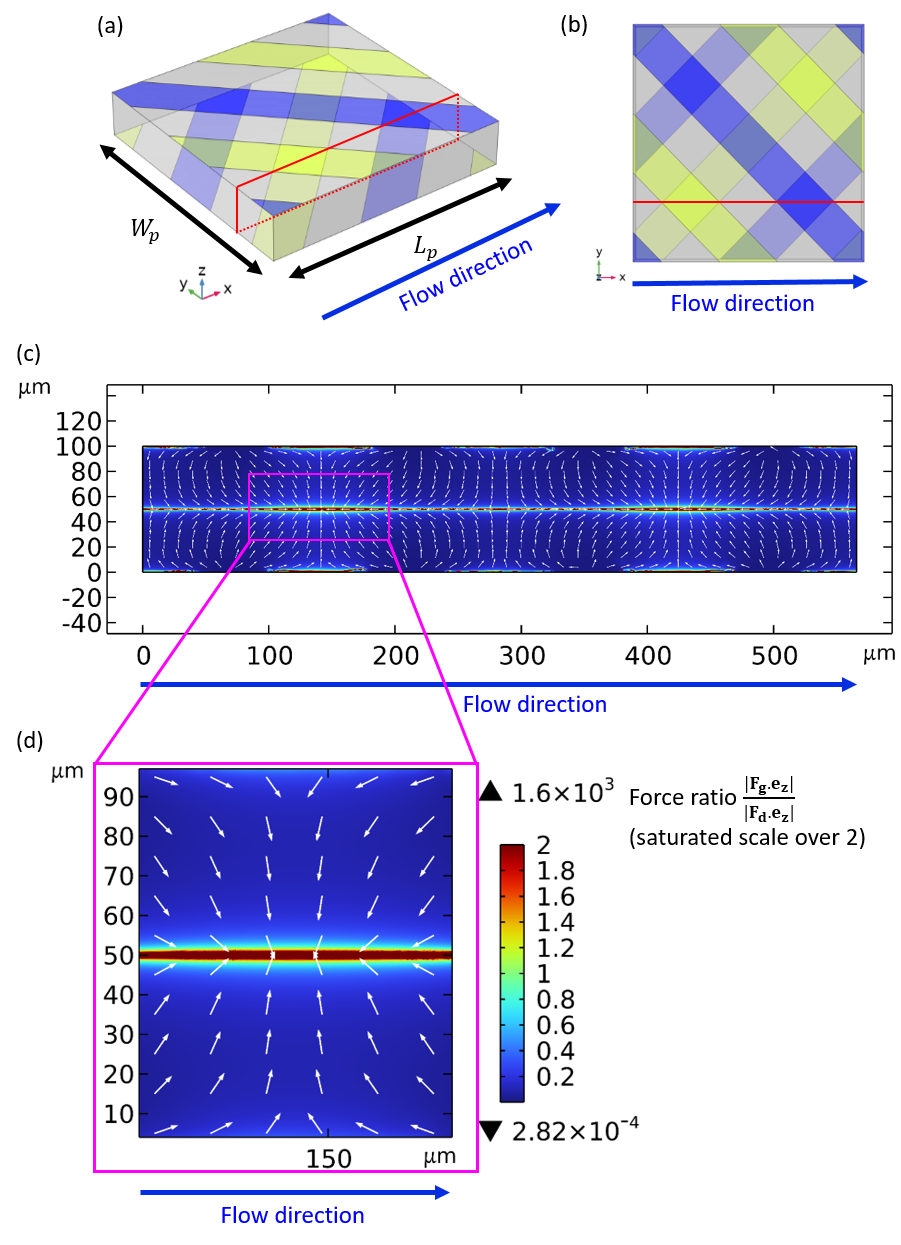}
    \caption{Investigation of the effect of gravity. Panels (a) and (b) illustrate the considered plane (red lines), which includes the particle focusing lines (at $y=y_s=W_p/4$). Panels (c) and (d) show the vertical component $\Tilde{\xi}_z$ of the reduced DEP field in absolute value $|\Tilde{\xi}_z|$. White arrows indicate the (normalized) DEP force field $\mathbf{F_d}$ direction and orientation. Panel~(d) is a zoomed view of Panel~(c) in the low $E^2$ zones where $|\Tilde{\xi}_z|$ is the lowest.}
    \label{fig_sedim}
\end{figure}

If the (uniform) particle density $\rho_p$ differs from that of the surrounding medium $\rho$, its weight is not balanced by Archimedes' thrust (buoyancy). The resulting gravity-induced force is, for spherical particles,
\begin{equation}
    \mathbf{F_g} = \dfrac{4}{3} \pi R_p^3  \Delta \rho \ \mathbf{g}
\end{equation}
with $\Delta \rho = \rho_p - \rho $ and $\mathbf{g}=-g \mathbf{e_z}$ the gravity field.  
In order to evaluate its relative importance with respect to the DEP force, we consider the ratio
\begin{equation}
    \dfrac{|\mathbf{F_g}\cdot\mathbf{e_z}|}{|\mathbf{F_d}\cdot\mathbf{e_z}|}=\dfrac{2}{3} \ \dfrac{\Delta \rho }{\epsilon_m |\Re(K^*)|} \ \dfrac{g}{|\xi_z|}
    \; .
\end{equation}
Figure~\ref{fig_sedim} investigates this ratio in cut planes containing the focusing lines (at $y=y_s$, see panels~\ref{fig_sedim}(a,b)). Panel~\ref{fig_sedim}(c) shows that the force ratio $\dfrac{|\mathbf{F_g}\cdot\mathbf{e_z}|}{|\mathbf{F_d}\cdot\mathbf{e_z}|}$ is very small (inferior to $0.2$) almost everywhere in the cut plane, which means that the effect of gravity is negligible in this part of the system. Importantly, however, this ratio diverges near the focusing lines (at $z=H/2=50$~µm) where $|\xi_z|$ and consequently $|\mathbf{F_d}.\mathbf{e_z}|$ vanishes. The stable vertical positions next to the channel mid-plane are therefore slightly below the latter, where $\dfrac{|\mathbf{F_g}\cdot\mathbf{e_z}|}{|\mathbf{F_d}\cdot\mathbf{e_z}|} = 1$. For the setup considered in Fig.~\ref{fig_sedim}, they are found approximately $2.5$~µm below $H/2$ (see Panel \ref{fig_sedim}(d)). Overall, the present analysis justifies the fact that we neglect the effect of gravity on particle trajectories in the results reported in the main text.

\clearpage

\end{document}